# Adaptive Frequency-Regulation Demand Response Using Substation Solar Irradiance Measurement in High-PV Power Systems

Shutang You, *Member, IEEE*

***Abstract*— This letter proposes a distributed and adaptive demand response approach for primary frequency regulation in high-PV power systems. Using solar irradiance measurement at substations, the proposed approach allows accurate estimation of real-time system inertia of high-PV systems, thus facilitating the estimation of power imbalance after a contingency. The test results from the U.S. Electric Reliability Council of Texas (ERCOT) high-PV penetration models validated its effectiveness.**

***Index Terms*— demand response (DR), under-frequency load shedding (UFLS), inertia, the rate of change of frequency (ROCOF), solar irradiance sensor, photovoltaic.**

## I. Introduction

Power grids in many places see an increase of renewable generation penetration. As power grid reliability is a critical factor in operation and planning, the impact of renewable energy on grid reliability has been extensively studied. One of the major impacts of renewable generation on power grid is the frequency response. The frequency response should be carefully monitored by measurement systems to ensure the power grid has adequate frequency regulation capabilities. Existing adaptive frequency-regulation demand response or under-frequency load shedding (UFLS) schemes use the rate of the change of frequency (ROCOF) and a pre-determined system inertia constant to estimate the proportion of power imbalance over the system size so that the system frequency drop can be retained while the outage magnitude is minimized [1]. Ref. [2] first proposed using the first stage load shedding to update the estimation of real-time system inertia. In high-PV power grids, system inertia becomes more dynamic due to hourly change in solar irradiance profiles and generation commitments.

This letter presents a distributed adaptive demand response approach using only local measurements at substations. To consider the impact of real-time PV output, this approach adopts solar irradiance sensors and conventional sensors at substations, to aid the estimation of system inertia. The inertia result is then used to estimate power imbalance and calculate the amount of demand response to restrain frequency drop. The proposed approach was validated by future high-PV models of the U.S. Electric Reliability Council of Texas (ERCOT) system.

## II. Distributed and Adaptive Demand Response

The key factor that impacts real-time system inertia in high-PV systems is real-time PV output, since PV cannot provide inertia response due to the absence of kinetic energy. The inertia of a system is comprised of the inertia of generators and loads, which can be represented as (1).

$$H_{System} = H_{G0} \cdot (1 - \mu_{PV}) + H_{L0} \tag{1}$$

where $H_{System}$ is the system per unit inertia constant. $H_{G0}$ and $H_{L0}$ are average per unit inertia constants of conventional generators and load, respectively. For a specific system, these two constants can be estimated fairly accurately [1]. $\mu_{PV}$ is the PV instantaneous penetration as a percentage. It can be obtained from (2).

$$\mu_{PV} = \sum_k \{\varphi(\bar{r}_{PV,k}) \bar{P}_{PV,k}\} / P_{System} \cdot \tag{2}$$

where $\bar{r}_{PV,k}$ is the solar irradiance (ten-minute moving average value to mitigate the impact of fast moving clouds). $\varphi(\,)$ is the function of per unit PV output versus solar irradiance. $\bar{P}_{PV,k}$ is the installed PV capacity. $k$ is the index of PV plants. $P_{System}$ is the system total load.

Since governor response is not significant due to deadbands during the initial frequency drop period, the system total power imbalance can be estimated using ROCOF.

$$P_{Imbalance} / P_{System} = 2 H_{System} \cdot ROCOF / f_N \tag{3}$$

where $P_{Imbalance}$ is the estimated power imbalance. $f_N$ is the nominal frequency. Substituting (1) and (2) into (3), one can obtain:

$$P_{Imbalance} = 2\frac{ROCOF}{f_N} \cdot \{H_{G0} + H_{L0}\} \cdot P_{System} - 2\frac{ROCOF}{f_N} \cdot H_{G0} \cdot \sum_k \{\varphi(\bar{r}_{PV,k}) \bar{P}_{PV,k}\} \tag{4}$$

The system PV output ($\sum\{\varphi(\bar{r}_{PV})\bar{P}_{PV}\}_k$) and system load ($P_{System}$) can be estimated by (5) using local measurements at each substation with frequency-regulation demand response.

$$\sum_k \{\varphi(\bar{r}_{PV,k}) \bar{P}_{PV,k}\} = \varphi(\bar{r}_{PV,i}) \cdot \sum_k \bar{P}_{PV,k} + e_{PV,i} \tag{5.a}$$

$$P_{System} = g(P_{SubTotalLoad,i}) + e_{S,\mathrm{i}} \tag{5.b}$$

where $\bar{r}_{PV,i}$ is the moving average of solar irradiance measured at substation $i$. (5.a) demonstrates that system PV output is estimated using only local solar irradiance measurements and the system installed capacity. $g(\,)$ is the mapping function between substation actual total load and the system total load. A simple mapping function can be a gain coefficient

This work also made use of the Engineering Research Center Shared Facilities supported by the Engineering Research Center Program of the National Science Foundation and DOE under NSF Award Number EEC-1041877 and the CURENT Industry Partnership Program.

Shutang You is with Department of Electrical Engineering and Computer Science, the University of Tennessee, Knoxville, TN, 37996, USA. (E-mail: syou3@utk.edu).

representing the proportion of substation load to system load. $e_{PV,i}$ and $e_{S,i}$ are the PV output estimation and system load mapping errors, using the load and irradiance measurements, respectively at substation $i$. Then the demand response amount in each substation can be obtained by (6).

$$P_{DR,i} = \rho_i 2\frac{ROCOF}{f_N} \cdot \left\{\{H_{G0} + H_{L0}\} \cdot g\left(P_{SubTotalLoad,i}\right) - H_{G0} \cdot \varphi\left(\bar{r}_{PV,i}\right) \cdot \sum_k \bar{P}_{PV,k}\right\} \quad (6)$$

where $\rho_i$ is the proportion of the demand response in substation $i$. Since the feeders connected to the substation may have distributed PV installed, the total load of the substation can be estimated as:

$$P_{SubTotalLoad,i} = P_{SubNetLoad,i} + \bar{P}_{PV,i} \cdot \varphi\left(\bar{r}_{PV,i}\right) \quad (7)$$

where $P_{SubTotalLoad,i}$ is the actual total load at substation $i$. $P_{SubNetLoad,i}$ is the measured load of substation $i$ considering the load offset by PV.

Since the parameters in (6) are quite stable (mainly determined by system characteristics) except for $ROCOF$, substation netload $P_{SubNetLoad,i}$, and substation solar radiance $\bar{r}_{PV,i}$, the demand response $P_{DR,i}$ can be estimated and predetermined using local measurements from each substation. The error of the total demand response $E_{P_{DR}}$ is the sum of mapping error of system load and system PV output at all substations weighted by the demand response proportion $\rho_i$.

$$E_{P_{DR}} = \sum_i\left\{e_{P_{DR,i}}\right\} = 2\frac{ROCOF}{f_N}\left\{\{H_{G0} + H_{L0}\}\sum_i\left(\rho_i \cdot e_{S,i}\right) - H_{G0} \cdot \sum_i\left(\rho_i \cdot e_{PV,i}\right)\right\} \quad (8)$$

Although the practical determination of $\rho_i$ will involve many factors such as load priority, utilities' practice on determining $\rho_i$ usually results to a small $E_{P_{DR}}$ value due to two reasons. a) Substations with more load and feeders are usually assigned with a larger $\rho_i$ due to their demand response capability. In fact, load profiles of these substations are more consistent with the system load profile. b) Due to the geographic correlation between load, population, rooftop PV capacity, and electricity prices (related to profits of utility-scale PV), assigning higher weights to the solar irradiance of substations with higher load benefits the estimation of real-time system average PV output.

Since the proposed scheme automatically considers the impact of spatial diversity of solar irradiance on the demand response amount, the summation of the activated demand response at all substations, which is the primary influence factor of frequency regulation performance, will be close to the imbalance amount. An additional advantage of the proposed method over conventional methods is that the result of the proposed method is the actual value of demand response amount, instead of the proportion of demand response at all load buses. The actual imbalance value facilitates the control of frequency support devices and provide flexibility to reallocate demand response amount when not all substations are equipped with demand response. Fig. 2 summarizes the scheme of the proposed approach.

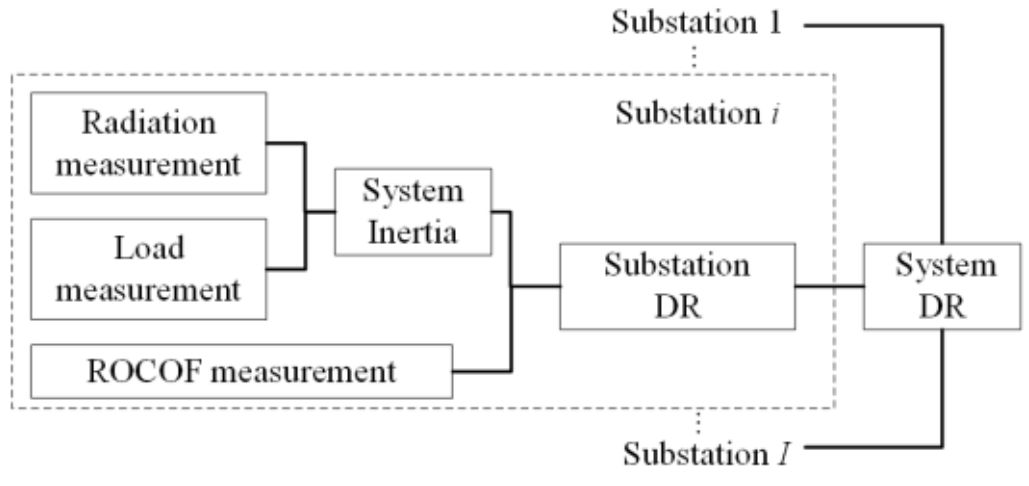

Fig. 2. Distributed and adaptive demand response for primary frequency regulation using substation solar irradiance measurement

## III. Case Studies

The proposed approach is applied to a set of models representing the U.S. ERCOT system (6,102 buses). This model has 15% wind power penetration (wind output is assumed to be the same for all scenarios) and various levels of instantaneous PV penetration. The threshold to activate demand response is 59.3 Hz with 40 cycles time delay including breaker operation. A generation trip contingency with 2,740 MW loss was simulated under multiple PV penetration scenarios. For conventional UFLS, 45% PV penetration is treated as the benchmark scenario to provide inertia information. The ERCOT system frequency using the proposed approach and the conventional adaptive UFLS method are shown in Fig. 1.

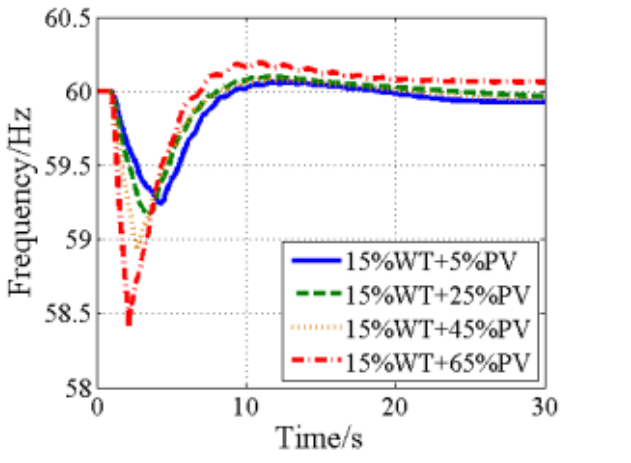

(a) Proposed approach

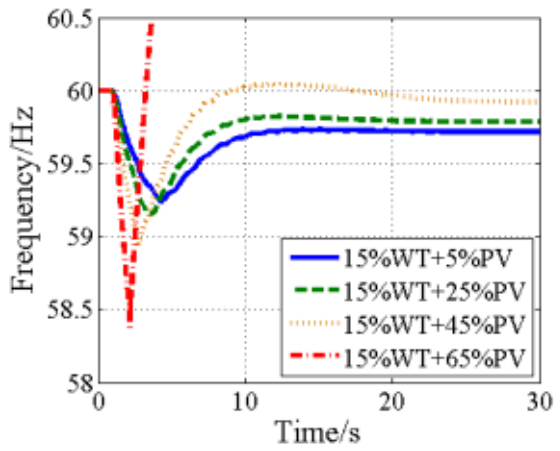

(b) Conventional approach

Fig. 1. Frequency comparison using proposed and conventional approaches

Fig. 1 shows obvious over shedding or under shedding by conventional UFLS. This error of conventional UFLS is due to absence of situational awareness on the PV output, thus over/underestimating the real-time system inertia and power imbalance. In contrast, the proposed distributed adaptive demand response method significantly improves estimation accuracy of the power imbalance by incorporating local solar irradiance measurements.

## IV. Conclusion

This letter proposed an improved adaptive demand response approach for primary frequency regulation using solar irradiance measurements at substations. The proposed method can improve the estimation accuracy of total power imbalance by considering real-time system PV output in a distributed manner. Adapted for highly variable high-PV systems, the proposed method does not require wide-area measurement and communication, overcoming the drawbacks (additional costs, time delay and reliability) of conventional adaptive UFLS.